# The Plane-Width of Graphs

Marcin Kamiński
Université Libre de Bruxelles
Marcin.Kaminski@ulb.be.ac

Paul Medvedev
University of Toronto[*]
pashadag@cs.toronto.edu

Martin Milanič
Universität Bielefeld
mmilanic@cebitec.uni-bielefeld.de

October 30, 2018

**Abstract**

Map vertices of a graph to (not necessarily distinct) points of the plane so that two adjacent vertices are mapped at least a unit distance apart. The *plane-width* of a graph is the minimum diameter of the image of the vertex set over all such mappings. We establish a relation between the plane-width of a graph and its chromatic number, and connect it to other well-known areas, including the circular chromatic number and the problem of packing unit discs in the plane. We also investigate how plane-width behaves under various operations, such as homomorphism, disjoint union, complement, and the Cartesian product.

## 1 Introduction

Given a simple, undirected, finite graph $G = (V, E)$, a *realization* of $G$ is a function $r$ assigning to each vertex a point in the plane such that for each $\{u, v\} \in E$, $d(r(u), r(v)) \geq 1$, where $d$ is the Euclidean distance. The *width* of a realization is the maximum distance between the images of any two vertices. In this paper, we introduce a new graph invariant, called the **plane-width** and denoted by $\mathrm{pw}(G)$, which is the minimum width of all realizations of $G$.

The plane-width of an edgeless graph is 0. To avoid trivialities, we only consider graphs with at least one edge. A realization of $G$ whose width equals $\mathrm{pw}(G)$ is called *optimal*. The image of the vertex set through a realization is called an *arrangement*. An edge $\{u,v\}$ in a graph will also be denoted by $uv$. Given a graph $G$, we denote with $V(G)$ its vertex set and with $E(G)$ its edge set. For terminology not defined here, we refer the reader to [9].

---

[*]Most of the work was done while on leave at the Universität Bielefeld.

**Related work.** A similar notion to the width of a realization is that of the *dilation coefficient*, defined by Pisanski and Žitnik in [19] as the ratio of the longest to the shortest edge length in a realization of a graph. In fact, since in every optimal realization the shortest length of an edge is precisely 1, the plane-width of a graph could be defined equivalently as the minimum ratio of the largest distance between two points of an arrangement and the shortest length of an edge. Notice that for complete graphs, the notion of plane-width coincides with the minimum possible dilation coefficient.

Belk and Connelly considered a more restricted notion, called *w-valid realizations*, where, for a function $w$ on the edges, each edge $\{u,v\} \in E$ imposes a constraint of the form $d(r(u), r(v)) = w(\{u,v\})$ [3]. The authors were concerned with necessary and sufficient conditions for a graph to have a $w$-valid realization for every reasonable choice of $w$.

## 2 Plane-width of complete graphs and odd wheels

The problem of determining the plane-width of complete graphs $K_n$ has previously appeared in the literature in different contexts: finding the minimum diameter of a set of $n$ points in the plane such that each pair of points is at distance at least one [5], or packing non-overlapping unit discs in the plane so as to minimize the maximum distance between any two disc centers [20]. A similar well-studied problem is that of computing the smallest diameter of a circle enclosing $n$ circles of unit diameter [13]. In this section, we review what is known about the plane-width of complete graphs and add our own results.

### Asymptotic behavior

The asymptotic behavior of $\mathrm{pw}(K_n)$ is largely determined. A lower bound is provided by the following result by Bezdek and Fodor.

**Lemma 2.1** ([5]). *For every $n$, $\mathrm{pw}(K_n) \geq \sqrt{\frac{2\sqrt{3}}{\pi}n} - 1$.*

An upper bound can be obtained by mapping vertices of $K_n$ to points of the triangular lattice such that they are contained in the smallest possible circle.

| $n$ | 2 | 3 | 4 | 5 | 6 | 7 | 8 |
|---|---|---|---|---|---|---|---|
| $\mathrm{pw}(K_n)$ | 1 | 1 | $\sqrt{2}$ | $\frac{1+\sqrt{5}}{2}$ | $2\sin 72°$ | 2 | $(2\sin(\pi/14))^{-1}$ |
| $\approx$ | 1 | 1 | 1.414 | 1.618 | 1.902 | 2 | 2.246 |

Table 1: Known values of $\mathrm{pw}(K_n)$ (in the last row rounded to three decimal places).

**Lemma 2.2.** *There exists a constant $C > 0$ such that for every $n \geq 2$,*

$$\mathrm{pw}(K_n) \leq \sqrt{\frac{2\sqrt{3}}{\pi}n} + C.$$

Together, these two bounds lead to the exact expression for the asymptotic behavior of $\mathrm{pw}(K_n)$.

**Theorem 2.3** ([2, 5, 12]).

$$\lim_{n\to\infty} \frac{\mathrm{pw}(K_n)}{\sqrt{n}} = \sqrt{\frac{2\sqrt{3}}{\pi}} \approx 1.05.$$

Interestingly, it was conjectured by Erdős and proved by Schürmann [20] that for all sufficiently large $n$, the optimal value of $\mathrm{pw}(K_n)$ is not attained by any lattice arrangement.

### Small complete graphs

The exact values of $\mathrm{pw}(K_n)$ are known only for complete graphs on at most 8 vertices. Clearly, $\mathrm{pw}(K_2) = \mathrm{pw}(K_3) = 1$. Below we compute $\mathrm{pw}(K_4)$ and $\mathrm{pw}(K_5)$ and report other known values (which are also grouped in Table 1).

**Proposition 2.4.**

(a) $\mathrm{pw}(K_4) = \sqrt{2}$ *and the unique optimal arrangement for $K_4$ is given by the corners of the unit square,*

(b) $\mathrm{pw}(K_5) = \frac{1+\sqrt{5}}{2}$ *and the unique optimal arrangement for $K_5$ is given by the corners of a regular pentagon with side length 1.*

*Proof.* It is easy to verify that the above arrangements have the desired diameters, so what remains to show is a matching lower bound on the plane-width and a proof of uniqueness.

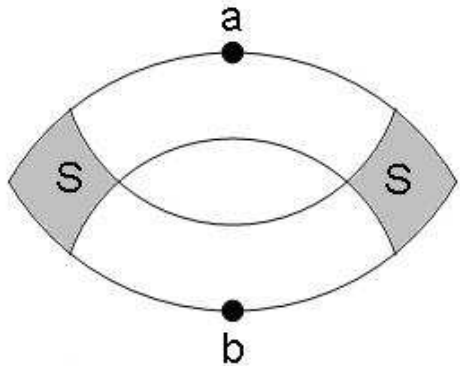

Figure 1: The situation in the proof of Proposition 2.4. The shaded area is the set $S$.

Suppose that a complete graph has an arrangement of width $d < (1+\sqrt{5})/2$ and let two points at distance $d$ in this arrangement be called $a$ and $b$. The remaining points of the arrangement must lie within the set $S$, the set of all the points at distance at least 1 and at most $d$ from both $a$ and $b$. The set $S$ is composed of two connected parts, each of diameter $d - 1/d < 1$ (see Figure 1). Therefore, each of the two parts can contain at most one vertex, and so the graph contains at most four vertices. This immediately gives us that $pw(K_5) \geq (1+\sqrt{5})/2$. Moreover, for $d < \sqrt{2}$, the shortest distance between any two points lying in different parts of $S$ is $\sqrt{4-d^2} > \sqrt{2}$. Hence, the two parts of $S$ cannot both contain a vertex, and the graph cannot contain more than 3 vertices. This gives us that $pw(K_4) \geq \sqrt{2}$.

To show uniqueness for $K_4$, note that for $d = \sqrt{2}$ there exists only one pair of points lying in different parts of $S$ such that the distance between them is not more than $\sqrt{2}$. For $K_5$, observe that for $d = (1+\sqrt{5})/2$ there exists only one pair of points within a connected part of $S$ that are at least a unit distance apart. By mapping two of the vertices of $K_5$ to such a pair, we restrict the remaining vertex to exactly one location – the closest point in the other part of $S$. □

The plane-width of $K_6$ has been reported to be $pw(K_6) = 2\sin 72°$, the optimal arrangement consisting of the center and the vertices of a regular pentagon of circumradius 1 [2]. (However we are not aware of a simple proof of this fact.) Bateman and Erdős [2] showed that $pw(K_7) = 2$ and that the unique optimal arrangement consists of the center and the vertices of a regular hexagon of side length 1. Bezdek and Fodor [5] proved that $pw(K_8) = (2\sin(\pi/14))^{-1} \approx 2.246$ and that the convex hull of every optimal arrangement of $K_8$ is the regular heptagon with unit sides. The current best upper bound on the plane-width of $K_9$ is 2.584306 by Audet et al. [1].

## Odd wheels

An *odd wheel* is the graph obtained from an odd cycle by adding a new vertex adjacent to all vertices of the cycle. The smallest odd wheel is $K_4$. We now generalize the result for the plane-width of $K_4$ to arbitrary odd wheels.

**Proposition 2.5.** *The plane-width of every odd wheel is equal to $\sqrt{2}$.*

*Proof.* Let $G$ be an odd wheel. To show that $\mathrm{pw}(G) \leq \sqrt{2}$, consider a proper 4-coloring of $G$, and map (the vertices of) each color class to a different vertex of the unit square.

Suppose now that $\mathrm{pw}(G) = d < \sqrt{2}$ and consider an arrangement $\mathcal{A}$ of $G$ of width $d$. Let $v^*$ denote the vertex adjacent to the remaining vertices of $G$. Assume without loss of generality that $v^*$ is mapped to the origin. Then, all the other vertices must be mapped to points at distance at least 1 and at most $d$ from the origin. Moreover, we can assume that one of the points other than the origin lies on the $x$-axis, and all the other points in $\mathcal{A}$ lie in the first quadrant (otherwise we can rotate the arrangement around the origin). Now, let $P$ denote the point $(1,0)$ and let $Q$ denote the point in the first quadrant that is at distance $d$ from $P$ and at distance 1 from the origin. Furthermore, let $\ell$ denote the line through $P$ perpendicular to the line segment $PQ$, and let $\ell'$ denote the line parallel to $\ell$ passing through the point $Q$.

We now rotate the arrangement counter-clockwise so that it lies entirely on or above the line $\ell$, and so that at least one of the points lies on $\ell$. Then, all the points of the rotated arrangement (except $v^*$) belong to the part of the first quadrant between the lines $\ell$ and $\ell'$ and between the two concentric circles of respective radii 1 and $d$ centered at the origin. We denote this set by $S$. The line parallel to $PQ$ and tangent to the outer circle defines (together with $\ell$, $PQ$ and $\ell'$) a rectangle containing $S$, with side lengths $d$ and $d - \sqrt{1 - d^2/4} < d/2$.

To complete the proof, we will now show that no arrangement of an odd cycle can be entirely contained in an rectangle $R$ with side lengths $d$ and $d/2$ where $d < \sqrt{2}$. Suppose the converse and assume (without loss of generality) that $R$ is axis parallel with horizontal side of length $d$. Let $C$ be an odd cycle mapped to points within $R$, and let $v_1, \ldots, v_{2k+1}$ be the cyclic order of the vertices of $C$. Furthermore, let $(x_1, y_1), \ldots, (x_{2k+1}, y_{2k+1})$ denote the images of the corresponding vertices in such an arrangement. For each $i = 1, \ldots, 2k+1$, let $\Delta x_i = x_{i+1} - x_i$ and $\Delta y_i = y_{i+1} - y_i$ (indices taken modulo $2k+1$). Since $(\Delta y_i)^2 \leq d^2/4 < 1/2$ and $(\Delta x_i)^2 + (\Delta y_i)^2 \geq 1$, we conclude that $|\Delta x_i| > \sqrt{2}/2$ for all $i$. Since the cycle is odd, there exist two consecutive indices $j$ and $j+1$ such that $\Delta x_j$ and $\Delta x_{j+1}$ are

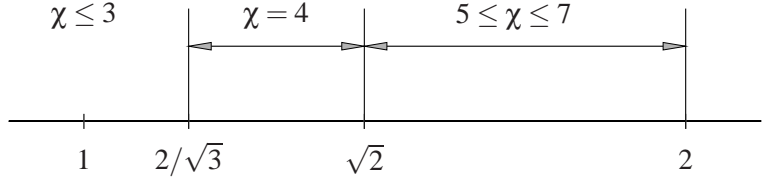

Figure 2: Relation between pw and $\chi$ for small values of these invariants.

of the same sign, say (without loss of generality) $\Delta x_j, \Delta x_{j+1} > 0$. However, this implies that $\Delta x_j, \Delta x_{j+1} > \sqrt{2}/2$ and therefore $x_{j+2} - x_j = \Delta x_j + \Delta x_{j+1} > \sqrt{2} > d$; a contradiction to the fact that the arrangement is entirely contained in $R$. $\square$

## 3 Plane-width and the chromatic number

In this section we establish a connection between the plane-width of a graph and its chromatic number.

**Graphs with small chromatic number**

For small values of the chromatic number, there is a strong relation between the plane-width of a graph and its chromatic number. The goal of this subsection is to prove the following theorem.

**Theorem 3.1.** *For all graphs G,*

(a) $\mathrm{pw}(G) = 1$ *if and only if* $\chi(G) \le 3$,

(b) $\mathrm{pw}(G) \notin (1, 2/\sqrt{3}]$,

(c) $\mathrm{pw}(G) \in (2/\sqrt{3}, \sqrt{2}]$ *if and only if* $\chi(G) = 4$,

(d) $\mathrm{pw}(G) \in (\sqrt{2}, 2]$ *if and only if* $\chi(G) \in \{5, 6, 7\}$.

In particular, every bipartite graph has plane-width exactly 1. Also, every graph with maximum degree 3, different from the complete graph on 4 vertices, has plane-width exactly 1. (By Brooks' Theorem such graphs are 3-colorable.) The plane-width of every planar graph is at most $\sqrt{2}$ (as such graphs are 4-colorable), and the plane-width of graphs embeddable on a torus is at most 2 (as such graphs are 7-colorable).

We start with some definitions. The term *plane set* will mean a set of points in the plane. Given a realization $r$ of a graph $G$ and a plane set $S$, we denote by $r^{-1}(S)$ the set $\{v \in V(G) : r(v) \in S\}$. The diameter of $S$ is defined as $\text{diam}(S) = \sup_{x,y \in S} d(x,y)$. We say that $S$ is $\delta$-*small* if $d(x,y) < \delta$ for all $x, y \in S$. Note that every set of diameter less than $\delta$ is $\delta$-small, but the converse does not hold; the diameter of a $\delta$-small set can be $\delta$. The importance of 1-small sets for the purposes of relating plane-width to coloring is based on the following statement, which follows directly from the definitions.

**Observation 3.2.** *Let $G$ be a graph and $r$ be a realization of $G$. For every 1-small plane set, the set $r^{-1}(S)$ is an independent set in $G$.*

The following lemma establishes a connection between the plane-width of a graph and its chromatic number.

**Lemma 3.3.** *Let $G$ be a graph and let $\delta = 1/\text{pw}(G)$. If every plane set of unit diameter can be partitioned into $k$ $\delta$-small sets, then $\chi(G) \leq k$.*

*Proof.* Let $G$ be a graph and let $\delta = 1/\text{pw}(G)$. Consider an arrangement $A$ of $G$ given by an optimal realization $r$. Then, $A$ is a plane set of diameter $\text{diam}(A) = \text{pw}(G)$. Suppose that every plane set of unit diameter can be partitioned into $k$ $\delta$-small sets. Then, for every $d > 0$, every plane set of diameter $d$ can be partitioned into $k$ $(\delta d)$-small sets. In particular, taking $d = \text{pw}(G)$, we can partition $A$ into $k$ 1-small sets $A_1, \ldots, A_k$. By Observation 3.2, each of the sets $r^{-1}(A_i)$ is an independent set in $G$.

Therefore, if every plane set of unit diameter can be partitioned into $k$ $\delta$-small sets, then the vertex set of $G$ can be partitioned into $k$ independent sets, which implies $\chi(G) \leq k$. □

This lemma gives us a method for translating upper bounds on $\text{pw}(G)$ into upper bounds on $\chi(G)$, which involves showing how to partition a plane set of unit diameter into sets of smaller diameter. We now apply this technique to graphs of small plane-width.

**Lemma 3.4.** *Every plane set of unit diameter can be partitioned in either of the following ways:*

(a) *Three ($\sqrt{3}/2$)-small sets,*

(b) *Four ($\sqrt{2}/2$)-small sets,*

(c) *Seven $(1/2)$-small sets.*

*Proof.* *(a)*. The proof was given by Boltjansky and Gohberg in [7], but for completeness we present it here. By a result of Pál [17], every plane set $S$ of unit diameter can be surrounded by a regular hexagon whose opposite sides are at unit distance. Having found such a hexagon $H$, we can cut it into three sets $H_1$, $H_2$ and $H_3$ of diameter $\sqrt{3}/2$, as follows: denoting the sides of the hexagon by $s_1, s_2, \ldots, s_6$ in a cyclic order, we cut the hexagon along the lines $l_1$, $l_2$, $l_3$ connecting the center of the hexagon with the midpoints of the sides $s_1$, $s_3$ and $s_5$ respectively. For each $i \in \{1,2,3\}$, we let $H_i$ be the subset of $H$ defined by the boundaries of $l_i$ and $l_{i+1}$, inclusive of $l_i$ and exclusive of $l_{i+1}$ (indices take modulo 3). Moreover, we assume that the center of the hexagon belongs to $H_1$ but not to $H_2$ and $H_3$ (see Figure 3a).

By construction, the sets $H_1$, $H_2$ and $H_3$ form a partition of $H$ and are each $(\sqrt{3}/2)$-small. Finally, the three $(\sqrt{3}/2)$-small sets that partition $S$ are given by $S_i = H_i \cap S$ for all $i \in \{1,2,3\}$.

*(b)*. Consider a plane set $S$ of unit diameter. $S$ is contained in a unit square. (Draw two lines parallel to the $y$-axis: through the left-most and the right-most point of the set, and draw another two lines parallel to the $x$-axis: through the top-most and bottom-most point of the set; take a unit square containing the region between the lines.)

Notice that there is a corner of the square not containing any point from $S$. (For any two endpoints of a diagonal the square, at most one can contain a point from $S$.) Without loss of generality we assume that the coordinates of such a corner are $(0,0)$, and the coordinates of other corners are $(0,1)$, $(1,0)$, $(1,1)$.

Draw two lines through the point $(1/2, 1/2)$: one parallel to the $x$-axis, one to $y$-axis. This divides the square into 4 smaller squares: NW, NE, SW, SE (see Figure 3b). To complete the proof, we remove some points to make all the small squares into $(\sqrt{2}/2)$-small sets: From NW remove $(0, 1/2)$ and $(1/2, 1/2)$, from NE remove $(1/2, 1/2)$ and $(1/2, 1)$, from SW remove $(0, 0)$ and $(1/2, 0)$, and from SE remove $(1/2, 1/2)$ and $(1, 1/2)$. These sets can be made pairwise disjoint by assigning each point that belongs to at least two sets in an arbitrary way to only one of the sets.

*(c)*. Again, we enclose the plane set $S$ of unit diameter in a regular hexagon $H$ whose opposite sides are at unit distance. Let us name the vertices of $H$ consecutively $p_0, \ldots, p_5$ and for $i = 0, 1, \ldots, 5$, let $m_i$ be the midpoint of edge $p_{i-1}p_i$ (indices taken modulo 6). Also, let $q_i$ be the point at distance $(\sqrt{3}-1)/2$ from $m_i$ on the line segment connecting $m_i$ and $m_{i+3}$. The convex hull of $q_0, \ldots, q_5$ is a hexagon, and let $R$ be the convex hull of $q_0, \ldots, q_5$, without the $q_i$'s. Notice that $R$ is $(1/2)$-small.

Let $R_i$ be the convex hull of $q_i, m_i, p_i, m_{i+1}, q_{i+1}$, without points $q_{i+1}$ and $m_{i+1}$. It is easy to verify that each of $R_i$'s is a $(1/2)$-small set. Therefore, $R, R_0, \ldots, R_5$ is a partition of $H$ into seven $(1/2)$-small sets (see Figure 3c). □

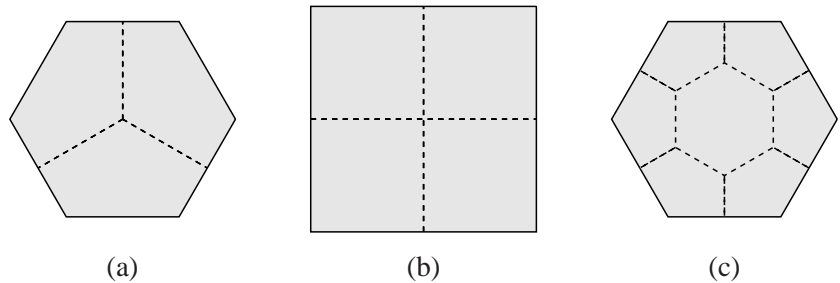

Figure 3: Partitions used in the proof of Lemma 3.4.

The following corollary is a direct consequence of Lemmas 3.3 and 3.4.

**Corollary 3.5.** *For any graph G,*

(a) *If* $\mathrm{pw}(G) \leq 2/\sqrt{3}$, *then* $\chi(G) \leq 3$,

(b) *If* $\mathrm{pw}(G) \leq \sqrt{2}$, *then* $\chi(G) \leq 4$,

(c) *If* $\mathrm{pw}(G) \leq 2$, *then* $\chi(G) \leq 7$.

Now we are ready to prove Theorem 3.1.

*Proof of Theorem 3.1. (a) and (b).* Any 3-colorable graph admits a realization of width 1 by assigning (the vertices of) each color class to a different vertex of the equilateral triangle with side length 1. On the other hand, if $\mathrm{pw}(G) \leq 2/\sqrt{3}$, then Corollary 3.5a gives that $\chi(G) \leq 3$. In turn, this implies that $\mathrm{pw}(G) \leq 1$.

*(c) and (d).* Observe that 4-colorable graphs admit a realization of width $\sqrt{2}$, by mapping (the vertices of) each color class to a different vertex of the unit square. Similarly, 7-colorable graphs admit a realization of width 2, by mapping each color class $C_1, \ldots, C_6$ to a different vertex of the regular hexagon $H$ of side length 1, and (the vertices of) the remaining color class to the center of $H$. Together with Corollary 3.5 these observations imply the theorem. □

## Graphs with large chromatic number

In this section we study the asymptotic behavior of $\mathrm{pw}(G)$ as $\chi(G) \to \infty$. We have already shown in Theorem 2.3 that $\mathrm{pw}(K_n) = \Theta(\sqrt{n})$. Now we prove, more generally, that the relation $\mathrm{pw}(G) = \Theta(\sqrt{\chi(G)})$ holds for arbitrary graphs as $\chi(G) \to \infty$.

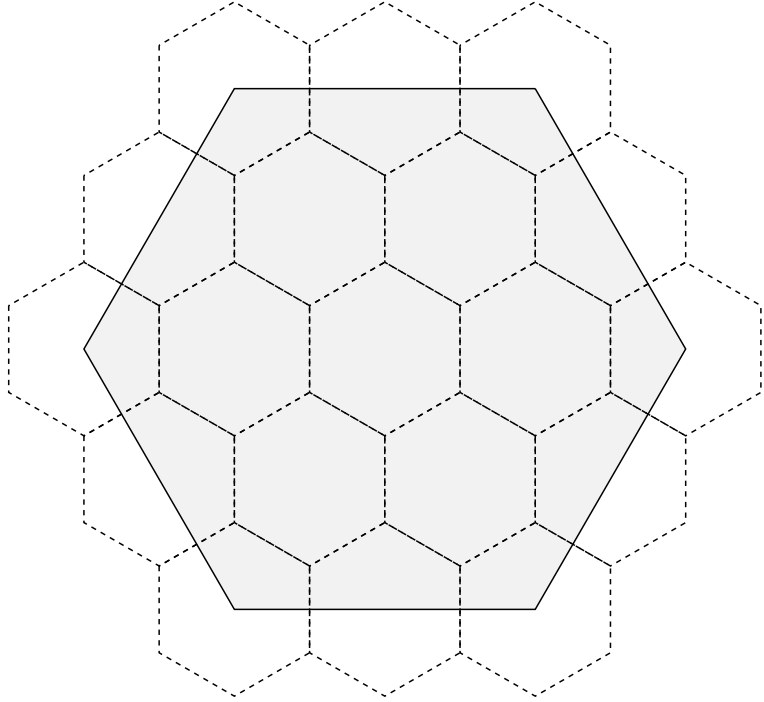

Figure 4: A partition of hexagon into 1-small sets ($6 \cdot \binom{2+1}{2} + 1 = 19$ hexagons) used in the proof of Lemma 3.6 for $t = 2$.

**Lemma 3.6.** *For every $\varepsilon > 0$ there exists an integer $k$ such that for all graphs $G$ of chromatic number at least $k$,*

$$\chi(G) < \left(\left(\frac{2}{\sqrt{3}} + \varepsilon\right) \cdot \mathrm{pw}(G)\right)^2.$$

*Proof.* Let $G$ be a graph, and consider an arrangement $A$ of $G$ given by an optimal realization of width $d = \mathrm{pw}(G)$. We can use the result of Pál [17] to enclose the arrangement in a regular hexagon $H$ whose opposite sides are at distance $d$. Let $t = \lceil 2d/3 \rceil$, and let $\mathcal{T}$ be a hexagonal tiling of the plane with hexagons of side length $d/(3t)$. $H$ (and with it $A$) can be translated and rotated so that $H$ is contained in the union of $6\binom{t+1}{2} + 1$ hexagons from $\mathcal{T}$. (First, rotate $H$ until it becomes parallel to the hexagons from $\mathcal{T}$, then translate it until each corner of $H$ coincides with the center of a hexagon in $\mathcal{T}$.) Moreover, by the choice of $t$, these $6\binom{t+1}{2} + 1 = 3t^2 +$ $3t+1$ hexagons can be turned into a collection of 1-small sets whose union contains $H$. This results in a proper coloring of $G$ with $3t^2 + 3t + 1$ colors. When $t$ is large enough, this expression can be bounded from above by $((2/\sqrt{3}+\varepsilon) \cdot \mathrm{pw}(G))^2$. □

**Lemma 3.7.** *For all graphs $G$, $\mathrm{pw}(G) \leq \mathrm{pw}(K_{\chi(G)})$.*

*Proof.* There is a bijection between color classes in an optimal coloring of $G$ and vertices of $K_{\chi(G)}$. The composition of such a bijection with a realization of $K_{\chi(G)}$ gives a realization of $G$. □

**Theorem 3.8.** *For every $\varepsilon > 0$ there exists an integer $k$ such that for all graphs $G$ of chromatic number at least $k$,*

$$\left(\frac{\sqrt{3}}{2}-\varepsilon\right)\sqrt{\chi(G)} < \mathrm{pw}(G) < \left(\sqrt{\frac{2\sqrt{3}}{\pi}}+\varepsilon\right)\sqrt{\chi(G)}.$$

*Proof.* Lemma 3.7 together with Theorem 2.3 give the upper bound. The lower bound follows from Lemma 3.6. □

Some questions regarding the plane-width of a graph can be answered via chromatic number by applying Theorem 3.8. For instance, the plane-width of almost every random graph (in the $G_{n,p}$ model with a fixed $p \in (0,1)$) is $\Theta(\sqrt{n/\log(n)})$ (since the chromatic number of almost every random graph is $\Theta(n/\log(n))$ [6]). Another example is the existence of graphs of arbitrarily large plane-width and girth (as there are graphs of arbitrarily large chromatic number and girth [11]).

## 4 Plane-width and circular chromatic number

An *r-circular realization* of a graph $G$ is a mapping which assigns each vertex of $G$ to a point on a circle of radius $r$ so that two adjacent vertices are mapped at distance at least 1 as measured along the circumference of the circle. The *circular chromatic number* of a graph $G$, denoted by $\chi_c(G)$, is defined as

$$\chi_c(G) = \inf\{\, 2\pi c \;:\; G \text{ admits a } c\text{-circular realization}\,\}.$$

Due to the fact that for all graphs $G$, $\chi(G) - 1 < \chi_c(G) \leq \chi(G)$, the circular chromatic number can be seen as a refinement of the chromatic number. The circular chromatic number is a well studied graph parameter (see [23] for a survey). In this section, we will establish a connection between the circular chromatic number and plane-width. This will allow us to apply some known results to obtain the following theorem, which should be viewed as complementing Theorem 3.1.

**Theorem 4.1.** *For every $\varepsilon > 0$ there exists*

(a) *A 4-chromatic graph G such that* $\mathrm{pw}(G) < 2/\sqrt{3} + \varepsilon$,

(b) *A 5-chromatic graph G such that* $\mathrm{pw}(G) < \sqrt{2} + \varepsilon$,

(c) *An 8-chromatic graph G such that* $\mathrm{pw}(G) < 2 + \varepsilon$.

We start out by using a graph's circular chromatic number to upper bound its plane-width.

**Lemma 4.2.** *For all graphs G*, $\mathrm{pw}(G) \leq \left[\sin\left(\frac{\pi}{\chi_c(G)}\right)\right]^{-1}$.

*Proof.* We can view any $r$-circular realization of $G$ as being defined by a function $f$ that maps each vertex to an angle $0 \leq \theta < 2\pi$. The location of vertex $v$ under this $r$-circular realization is then given in polar coordinates by $r$ and $f(v)$. For a pair of vertices $v$ and $w$, denote by $\Delta(v,w)$ the angle (in the range of $[0, \pi]$) between the images of these two vertices. Denote by $\chi_c$ the circular chromatic number of $G$. We know that there exists a $(\chi_c/2\pi)$-circular realization, defined by a function $f$. Moreover, the length of the arc between the images of any two adjacent vertices $v$ and $w$ must be at least one, meaning that $\Delta(v,w) \geq 2\pi/\chi_c$.

Now let $r = (2\sin(\pi/\chi_c))^{-1}$ and consider the $r$-circular realization defined by $f$. Using basic trigonometry, we get that the distance between the locations of any two vertices $v$ and $w$ is $2r\sin(\Delta(v,w)/2)$. Plugging in $r$, we verify that the distance between the locations of any two adjacent vertices $v$ and $w$ is at least one. Therefore this is a realization, and its width is at most $2r = (\sin(\pi/\chi_c))^{-1}$. □

Using the above connection, we are able to translate a theorem by Vince about the existence of graphs with arbitrary rational value of the circular chromatic number into a result about the existence of $k$-chromatic graphs with a bounded plane-width.

**Theorem 4.3** ([21]). *For every rational number $q \geq 2$, there exists a graph G with $\chi_c(G) = q$.*

**Lemma 4.4.** *For every $k \geq 3$ and every $\varepsilon > 0$, there exists a k-chromatic graph G such that* $\mathrm{pw}(G) < [\sin(\pi/(k-1))]^{-1} + \varepsilon$.

*Proof.* Fix $k \geq 3$ and $\varepsilon > 0$. Let $\delta \in (0,1)$ be a number such that

$$[\sin(\pi/(k-1+\delta))]^{-1} < [\sin(\pi/(k-1))]^{-1} + \varepsilon.$$

Furthermore, let $q$ be a rational number such that $q \in (k-1, k-1+\delta)$. By Theorem 4.3, there exists a graph $G$ of circular chromatic number $q$. Since $k-1 < q < k$ and $\chi(G) = \lceil \chi_c(G) \rceil$, we conclude that $\chi(G) = k$. To upper bound the plane-width of $G$, we use Lemma 4.2:

$$\text{pw}(G) \leq \frac{1}{\sin(\frac{\pi}{\chi_c(G)})} = \frac{1}{\sin(\frac{\pi}{q})} \leq \frac{1}{\sin(\frac{\pi}{k-1+\delta})} < \frac{1}{\sin(\frac{\pi}{k-1})} + \varepsilon.$$

The second inequality follows from the fact that the function $f : x \mapsto (\sin(\pi/x))$ is non-decreasing for $x \geq 2$. □

Notice that for large $\chi_c(G)$, the bound of Lemma 4.2 (and hence of Lemma 4.4) becomes very weak since it grows linearly in $\chi_c(G)$, whereas Theorem 3.8 tells us that $\text{pw}(G)$ grows as the square root of $\chi_c(G)$. However, for small $\chi_c(G)$, we can still get meaningful bounds. In particular, we can prove Theorem 4.1.

*Proof of Theorem 4.1.* Parts (a) and (b) are a direct consequence of Lemma 4.4. For part (c), though the Lemma also gives a bound, we can get a tighter one with the following argument.

Given $\varepsilon > 0$, let $C$ be a circle of diameter $2 + \varepsilon$. Let $n \geq 2$ be the smallest positive integer such that

$$(2+\varepsilon) \sin\left(\frac{n}{6n+1} \cdot \pi\right) \geq 1. \tag{1}$$

Consider a graph $G$ whose vertex set consists of $6n+1$ points $p_1, \ldots, p_{6n+1}$ spread equidistantly on $C$. Two vertices of $G$ are joined by an edge if and only if the Euclidean distance between the corresponding points is at least 1. It follows from equation (1) that two vertices $p_i$ and $p_j$ with $i < j$ are adjacent if and only if $\min\{j-i, 6n+1+i-j\} \geq n$. By a result of Vince [21], the circular chromatic number of $G$ is $(6n+1)/n = 6+1/n$, which implies that $\chi(G) = \lceil \chi_c(G) \rceil = 7$. Now, let $G^*$ be the graph whose vertex set consists of the center of $C$, together with the points on $C$ corresponding to the vertices of $G$. Moreover, let $E(G^*) = E(G) \cup E'$, where $E'$ denotes the set of edges connecting the center of $C$ to all other vertices. By construction, we have $\chi(G^*) = \chi(G) + 1 = 8$. Moreover, the defining collection of points gives an arrangement of $G^*$ of width less than $2+\varepsilon$.

□

## 5 Plane-width and graph operations

In this section we study how different graph operations change the plane-width.

### Homomorphisms and perfect graphs

A *homomorphism* of a graph $G$ to a graph $H$ is an adjacency-preserving mapping, that is a mapping $\phi : V(G) \to V(H)$ such that $\{\phi(u), \phi(v)\} \in E(H)$ whenever $\{u, v\} \in E(G)$. We say that a graph $G$ is *homomorphic* to a graph $H$ if there exists a homomorphism of $G$ to $H$.

Any graph with chromatic number $\chi(G)$ is homomorphic to $K_{\chi(G)}$. The following lemma generalizes Lemma 3.7.

**Theorem 5.1.** *Let $G$ be a graph homomorphic to a graph $H$. Then, $\mathrm{pw}(G) \leq \mathrm{pw}(H)$.*

*Proof.* Let $G$ be a graph homomorphic to a graph $H$, and let $\phi : V(G) \to V(H)$ be a homomorphism of $G$ to $H$. Fix an optimal realization $r$ of $H$, and consider the realization $r'$ of $G$ given by $r'(v) = r(\phi(v))$ for each $v \in V(G)$. Then, $r'$ is a realization of $G$. This is because if $\{u, v\} \in E(G)$ then $d(r'(u), r'(v)) = d(r(\phi(u)), r(\phi(v))) \geq 1$. The inequality follows from the facts that $\{\phi(u), \phi(v)\} \in E(H)$ and that $r$ is a realization of $H$.

Moreover, since the set $\{r'(v) : v \in V(G)\} = \{r(\phi(v)) : v \in V(G)\}$ is a subset of the set $\{r(v) : v \in V(H)\}$, the width of $r'$ does not exceed that of $r$. In particular, this implies that $\mathrm{pw}(G) \leq \mathrm{pw}(H)$. □

We denote by $\omega(G)$ the maximum size of a clique in $G$.

**Corollary 5.2.**

(a) *For every graph $G$ and its subgraph $G'$, $\mathrm{pw}(G') \leq \mathrm{pw}(G)$.*

(b) *For every graph $G$, $\mathrm{pw}(G) \geq \mathrm{pw}(K_{\omega(G)})$.*

*Proof.* It is enough to observe that if $G'$ is a subgraph of $G$, then $G'$ is homomorphic to $G$. Part (b) follows from (a), since every graph $G$ contains $K_{\omega(G)}$ as a subgraph. □

These observations together with Lemma 3.7 imply that for graphs whose chromatic number coincides with their maximum clique size, the plane-width is a function of the chromatic number. In particular, this is the case for perfect graphs.

**Corollary 5.3.** *Let $G$ be a graph such that $\chi(G) = \omega(G)$. Then, $\mathrm{pw}(G) = \mathrm{pw}(K_{\chi(G)})$. In particular, if $G$ is a perfect graph, then $\mathrm{pw}(G) = \mathrm{pw}(K_{\chi(G)})$.*

### Join of graphs and edge subdivision

Given two graphs $H_1$ and $H_2$, let $H_1 \oplus H_2$ denote the graph (the *join* of $H_1$ and $H_2$) obtained from $H_1$ and $H_2$ by making every vertex of $H_1$ adjacent to every vertex of $H_2$.

**Theorem 5.4.** *For every two graphs $H_1$ and $H_2$,*

$$\mathrm{pw}(H_1 \oplus H_2) \leq \mathrm{pw}(H_1) + \mathrm{pw}(H_2) + 1.$$

*Proof.* For $i = 1, 2$, let $S_i$ be an arrangement of $H_i$, and let $a_i$, $b_i$ be two points of $S_i$ at distance $\mathrm{pw}(H_i)$. Place $S_1$ and $S_2$ in such a way that $b_1, a_1, a_2, b_2$ are collinear and placed on the line $\ell$ in this order, with $a_1, a_2$ being at distance 1.

We will show that this is an arrangement of $H_1 \oplus H_2$. Let $\ell_i$ be the line perpendicular to $\ell$ and passing through $a_i$, for $i = 1, 2$. Lines $\ell_1, \ell_2$ divide the plane into three parts. Notice that the part not containing $b_1$ or $b_2$ does not contain any point of $S_1$ or $S_2$, respectively. If it did, the distance between that point and $b_1$ (or $b_2$) would be greater than the diameter of $S_1$ (or $S_2$).

Now we will show that the diameter of $S_1 \cup S_2$ is at most $\mathrm{pw}(H_1) + \mathrm{pw}(H_2) + 1$. Consider two points $x_i \in S_i$, for $i = 1, 2$. From triangle $x_1, a_1, a_2$, the distance between $x_1$ and $a_2$ should be at most $\mathrm{pw}(H_1) + 1$. Now from the triangle $x_1, x_2, a_2$, the distance between $x_1$ and $x_2$ should be at most $\mathrm{pw}(H_1) + \mathrm{pw}(H_2) + 1$. □

The following corollary is a consequence of Theorem 5.4 and the fact that $G \subseteq (G - H) \oplus H$, where $G - H$ is the subgraph of $G$ induced by $V(G) \setminus V(H)$.

**Corollary 5.5.** *For every graph $G$ and its induced subgraph $H$,*

$$\mathrm{pw}(G) \leq \mathrm{pw}(G - H) + \mathrm{pw}(H) + 1.$$

We also consider the operation of doubly subdividing an edge of a graph, where some edge $uv$ is removed, two new vertices $x$ and $y$ are added, and finally the edges $ux, xy, yv$ are added.

**Corollary 5.6.** *Let $G$ be a graph and $G'$ the graph obtained from $G$ by doubly subdividing an edge. Then,*

$$\mathrm{pw}(G) - 1 \leq \mathrm{pw}(G') \leq \mathrm{pw}(G).$$

*Proof.* Notice that $G'$ is homomorphic to $G$, so $\mathrm{pw}(G') \leq \mathrm{pw}(G)$. The other inequality follows by observing that the subgraph $H$ of $G$ induced by $V(G) - \{u\}$, where $u$ is an endpoint of the subdivided edge, is a subgraph of $G'$. Therefore, $\mathrm{pw}(G) \leq \mathrm{pw}(H) + 1 \leq \mathrm{pw}(G') + 1$. The first inequality follows by Corollary 5.5, and the second one by Corollary 5.2. □

Notice that doubly subdividing *each* edge of a graph results in a 3-colorable graph, therefore in a graph of plane-width 1. Furthermore, if $G'$ is the graph obtained from a graph $G$ by doubly subdividing the edge $e$, and $G'' = G' + e$, then $\text{pw}(G'') = \text{pw}(G)$. In particular, when computing the plane-width of a graph, we can delete from the graph every pair of adjacent vertices of degree 2 contained in a four-cycle. This approach can be generalized to the case when a graph $G$ contains a bipartite graph $H$ that is attached to the rest only through one of its edges, say $\{x,y\}$. In this case, $\text{pw}(G) = \text{pw}(G - (V(H)\setminus\{x,y\}))$.

## Cartesian products

The *Cartesian product* of two graphs $G$ and $H$ is the graph $G \square H$ with vertex set $V(G) \times V(H)$ and edge set $\{(u,x)(v,y) : (u,x),(v,y) \in V(G) \times V(H), u = v$ and $xy \in E(H)$ or $x = y$ and $uv \in E(G)\}$. Since both $G$ and $H$ are subgraphs of $G \square H$, Corollary 5.2 implies that $\text{pw}(G \square H) \geq \max\{\text{pw}(G), \text{pw}(H)\}$. In the following lemma we provide an exact and an asymptotic upper bound.

**Theorem 5.7.**

(a) For every two graphs $G$ and $H$,
$$\text{pw}(G \square H) \leq \text{pw}(G) + \text{pw}(H).$$

(b) For every $\varepsilon > 0$ there exists a $p > 0$ such that for every two graphs $G$ and $H$ of plane-width at least $p$,
$$\text{pw}(G \square H) \leq \left(\sqrt{\frac{8}{\sqrt{3}\pi}} + \varepsilon\right) \max\{\text{pw}(G), \text{pw}(H)\}.$$

*Proof.* To see the first inequality, fix a pair $r_G$ and $r_H$ of optimal realizations of $G$ and $H$, and consider a mapping $r$ of the vertices of $G \square H$ to the plane, given by $r((u,x)) = r_G(u) + r_H(x)$, for every $(u,x) \in V(G \square H)$. Let us verify that $r$ is a realization of $G \square H$. We only need show that $d(r((u,x)), r((u,y))) \geq 1$ whenever $\{x,y\} \in E(H)$; the other type of required inequalities will follow by analogy. So let $\{x,y\} \in E(H)$. Then $d(r((u,x)), r((u,y))) = d(r_G(u) + r_H(x), r_G(u) + r_H(y)) = d(r_H(x), r_H(y)) \geq 1$; the inequality follows since $r_H$ is a realization of $H$. Finally, the fact that the width of $r$ does not exceed the sum of the widths of $r_G$ and $r_H$ is an easy consequence of the triangle inequality: Let $(u,x),(v,y) \in V(G \square H)$. Then $d(r((u,x)), r((v,y))) \leq d(r((u,x)), r((v,x))) + d(r((v,x)), r((v,y))) = d(r_G(u), r_G(v)) + d(r_H(x), r_H(y)) \leq \text{pw}(G) + \text{pw}(H)$.

The inequality in (b) is a direct consequence of Theorem 3.8 and the fact that $\chi(G \square H) = \max\{\chi(G), \chi(H)\}$ [22]. □

**Disjoint union**

Given two graphs $G$ and $H$, we denote by $G \uplus H$ the disjoint union of $G$ and $H$. Again, Corollary 5.2 yields the inequality $\mathrm{pw}(G \uplus H) \geq \max\{\mathrm{pw}(G), \mathrm{pw}(H)\}$. We now give an upper bound.

**Theorem 5.8.** *For every two graphs $G$ and $H$, we have that*

$$\mathrm{pw}(G \uplus H) \leq \max\left(\mathrm{pw}(G), \mathrm{pw}(H), \frac{1}{\sqrt{3}}(\mathrm{pw}(G) + \mathrm{pw}(H))\right).$$

*Proof.* We use the result of Pál [17] that every plane set of diameter $d$ can be enclosed in a regular hexagon whose opposite sides are at a distance of $d$ apart. We enclose some optimal arrangements of $G$ and $H$ in regular hexagons with opposite sides at a distance of $\mathrm{pw}(G)$ and $\mathrm{pw}(H)$, respectively. We center both hexagons (and their corresponding arrangements) at the origin, and we rotate one of the hexagons so that its edges are parallel to the other one. In this arrangement, the maximum distance is achieved by either two points from $G$, two points from $H$, or from one point in $G$ and one point in $H$. For the last case, this distance is maximized by two points in opposite corners of their respective hexagons, with the distance being the sum of the halves of the diameters of the hexagons. □

**Complement of a graph**

There are two known results relating the chromatic number of a graph $G$ and the chromatic number of its complement co-$G$.

**Theorem 5.9** ([4], pp. 330 – 332). *For any graph $G$ on $n$ vertices,*

(a) $\chi(G) + \chi(co\text{-}G) \leq n + 1$,

(b) $\chi(G) \cdot \chi(co\text{-}G) \leq \left(\frac{n+1}{2}\right)^2$.

These two bounds can be combined to deduce the inequality $\sqrt{\chi(G)} + \sqrt{\chi(co\text{-}G)} \leq \sqrt{2(n+1)}$. Combining this with Lemma 2.2 and Lemma 3.7, we obtain the following results.

**Theorem 5.10.**

(a) *For every $\varepsilon > 0$ there exists an integer $N$ such that for all graphs $G$ with $n > N$ vertices,*

$$\mathrm{pw}(G) \cdot \mathrm{pw}(co\text{-}G) \leq \left(\frac{\sqrt{3}}{\pi} + \varepsilon\right)(n+1).$$

(b) There exists a constant $C > 0$ such that for all graphs $G$ on $n$ vertices,

$$\text{pw}(G) + \text{pw}(\text{co-}G) \leq \left(2\sqrt{\frac{\sqrt{3}}{\pi}}\right)\sqrt{n} + C.$$

In particular, if $G$ is self-complementary, then $\text{pw}(G) \leq \left(\sqrt{\frac{\sqrt{3}}{\pi}}\right)\sqrt{n} + C$.

## 6 Generalizations

**Other norms**

One possible generalization of the plane-width is to consider distance measures different from the Euclidean norm. If the plane is equipped with the $\ell_p$ norm, for some $1 \leq p \leq \infty$, we denote the corresponding plane-width of the graph by $\text{pw}_p(G)$. Using similar techniques as for the proof of Theorem 3.8, we could prove its more general version.

**Theorem 6.1.**

(a) *There exist constants $0 < c_1 < c_2$ such that for every $p > 1$ and for every $\varepsilon > 0$ there exists an integer $k$ such that for all graphs $G$ of chromatic number at least $k$,*

$$(c_1 - \varepsilon)^{1/p} \sqrt{\chi(G)} < \text{pw}_p(G) < (c_2 + \varepsilon)^{1/p} \sqrt{\chi(G)}.$$

(b) *For every graph $G$,*

$$\sqrt{\chi(G)} - 1 \leq \text{pw}_\infty(G) < \sqrt{\chi(G)}.$$

**Other dimensions**

We can also consider realizations of graphs in higher dimensions. When the realization maps vertices of the graph to $R^d$, the corresponding version of the plane-width is denoted by $\text{pw}^{(d)}(G)$. It is easy to see that $\text{pw}^{(1)}(G) = \chi(G) - 1$; hence we can view $\text{pw}^{(d)}(G)$ as a multi-dimensional generalization of the chromatic number. In the case of 2 dimensions, as proved in Theorem 3.8, $\text{pw}(G) = \Theta(\sqrt{\chi(G)})$. One can show that for $d$ dimensions, $\text{pw}^{(d)}(G) = \Theta(\chi(G)^{1/d})$.

In 1932, Borsuk presented the following conjecture.

**Borsuk's Conjecture** ([8]). *For every $d \geq 1$, every convex body in $\mathbb{R}^d$ can be partitioned into $d+1$ sets of smaller diameter.*

An ingenious construction of Kahn and Kalai [15] disproved this conjecture for $d = 1325$ and every $d \geq 2015$. Subsequent efforts showed that the conjecture fails for all $d \geq 298$ (e.g. [14]). However, the conjecture was proved to be true for $d = 2$ [7, 8] and also for $d = 3$ [10, 18]. From the fact that the Borsuk's Conjecture is true for the 3-dimensional space, we obtain the following theorem (similar to Theorem 3.1a).

**Theorem 6.2.** *For all graphs G,* $\mathrm{pw}^{(3)}(G) \leq 1$ *if and only if* $\chi(G) \leq 4$.

In general, the technique of partitioning sets of unit diameter into $k$ $\delta$-small sets can be applied to three-dimensional point sets to obtain results similar to Theorem 3.1 for $\mathrm{pw}^{(3)}$.

## Hypergraphs

We can also define the notion of plane-width for hypergraphs. (For definitions related to hypergraphs, see [4].) A *realization of a hypergraph* $\mathcal{H} = (V, \mathcal{E})$ is a function $r$ assigning to each vertex a point in the plane such that for each hyperedge $E \in \mathcal{E}$ of cardinality at least 2 there exists a pair of vertices $u, v \in E$ with $d(r(u), r(v)) \geq 1$, where $d$ is the Euclidean distance. The *width* of a realization is the maximum distance between the images of any two vertices and the **plane-width** of $\mathcal{H}$ denoted by $\mathrm{pw}(\mathcal{H})$, is the minimum width of all realizations of $\mathcal{H}$.

Recall that a *k-coloring* of a hypergraph $\mathcal{H}$ is a function which assigns to each vertex one of the colors $\{1, \ldots, k\}$ such that vertices of no hyperedge with more than 1 element receive the same color. The least integer $k$ for which $\mathcal{H}$ admits a $k$-coloring is called the chromatic number of $\mathcal{H}$ and denoted by $\chi(\mathcal{H})$.

First, let us proof a result similar to Lemma 3.7.

**Lemma 6.3.** *For all hypergraphs* $\mathcal{H}$, $\mathrm{pw}(\mathcal{H}) \leq \mathrm{pw}(K_{\chi(\mathcal{H})})$.

*Proof.* Fix a $\chi(\mathcal{H})$-coloring of $\mathcal{H}$ and an optimal arrangement for $K_{\chi(\mathcal{H})}$. For every vertex $v$ of $\mathcal{H}$, map $v$ to the point of the arrangement of $K_{\chi(\mathcal{H})}$ corresponding to $v$'s color. Notice that every hyperedge with at least two vertices contains a pair of vertices with different colors, and those will be mapped at distance are least 1 apart. □

All the upper bounds on the chromatic number in terms of the plane-width presented above use the same geometric technique. We divide an arrangement into a number of 1-small sets and then use the fact the vertices mapped to a given set are independent. The same technique could be used for hypergraphs.

Let $W$ be a subset of vertices of a hypergraph $\mathcal{H}$ which includes some hyperedge of $\mathcal{H}$ with at least 2 vertices (i.e., $W$ is not independent). There are at least two vertices belonging to this hyperedge (so also to $W$) that must be mapped at distance at least 1 apart. Hence, the image of $W$ is not a 1-small set.

The two observations together imply that some of our results can be reproved for hypergraphs. In particular, we have a theorem similar to Theorem 3.8.

**Theorem 6.4.** *For every $\varepsilon > 0$ there exists an integer $k$ such that for all hypegraphs $\mathcal{H}$ of chromatic number at least $k$,*

$$\left(\frac{\sqrt{3}}{2} - \varepsilon\right) \sqrt{\chi(\mathcal{H})} < \mathrm{pw}(\mathcal{H}) < \left(\sqrt{\frac{2\sqrt{3}}{\pi}} + \varepsilon\right) \sqrt{\chi(\mathcal{H})}.$$

# 7 Discussion and open problems

In this paper, we have introduced the plane-width of a graph and studied some of its basic properties, including its value for certain graphs, its relation to the chromatic number, and its behavior under certain graph operations. In addition to the possibility of tightening some of the bounds presented in this paper, there remain some deeper unanswered questions. We discuss some of them here.

**Relation to chromatic number**

We have seen that $\mathrm{pw}^{(1)}(G) = \chi(G) - 1$ (where instead of the plane we map vertices to a line), and so it is natural to view $\mathrm{pw}(G) = \mathrm{pw}^{(2)}(G)$ as an extension of the chromatic number to two dimensions. Though asymptotically we have shown that $\mathrm{pw}(G) = \Theta(\sqrt{\chi(G)})$, the connection on a finer scale remains unclear. Basic questions remain unanswered – for example, is it true for all graphs that $\mathrm{pw}(G) \leq \mathrm{pw}(H)$ if and only if $\chi(G) \leq \chi(H)$? Can two non-bipartite graphs have the same plane-width but different chromatic numbers? We pose the following more general problem.

**Problem 1.** *Let $\mathbb{P} = \{\mathrm{pw}(G) : G \text{ is a graph}\}$. Determine whether there exists a function (a monotone function) $f : \mathbb{P} \to \mathbb{Z}$ such that for every non-bipartite graph $G$, $f(\mathrm{pw}(G)) = \chi(G)$.*

The existence of such a function would imply that the chromatic number of a graph can be determined solely from its plane-width. Independently of the existence of such a function, however, we would like to know what the plane-width of a graph can tell us about its chromatic number. To this end, we pose the following question.

**Problem 2.** *For $k \geq 5$ such that $k \neq 7$, what is the value of*

$$\inf\{\mathrm{pw}(G) : \chi(G) = k\}?$$

We have shown in this paper that the corresponding value is 1 for $k \in \{2,3\}$, for $k = 4$, it is $2/\sqrt{3}$, and for $k = 7$, it is 2.

As pointed out in Section 2, the problem of determining the plane-width of complete graphs has appeared in the literature in different contexts (packing non-overlapping unit discs in the plane so as to minimize the maximum distance between any two disc centers, finding the minimum diameter of a well-spaced set of points in the plane). Complete graphs also play an important role in bounding the plane-width, since the bound $\mathrm{pw}(K_{\omega(G)}) \leq \mathrm{pw}(G) \leq \mathrm{pw}(K_{\chi(G)})$ is tight for some classes of graphs. Therefore, we think that the following subproblem of Problem 1 is important in its own right.

**Problem 3.** *Determine whether $\mathrm{pw}(K_n) < \mathrm{pw}(K_{n+1})$ holds for all $n \geq 3$.*

**Algorithmic aspects**

In this paper, we have not discussed any algorithms that would compute the plane-width of a particular graph $G$. However, the idea of finding a realization of small width is very natural, and can be used to model any problem where objects must be placed not too far away from each other, while maintaining some distance between certain pairs of objects.

The fact that determining if a graph is 3-colorable is NP-hard, along with Theorem 3.1, immediately shows that computing the plane-width, or approximating it within the factor of $2/\sqrt{3}$, is also NP-hard. If we are also willing to concede certain computation complexity assumptions, then we can use Theorem 3.8 to transfer the best known inapproximability result for chromatic number [16] to plane-width to get that plane-width is inapproximable in polynomial time within a factor of $O((n/2^{(\log n)^{3/4+\gamma}})^{1/2})$, for any $\gamma > 0$.

Besides finding general algorithms, we think it is of interest to focus on complete graphs, given their importance to determining bounds for plane-width. In particular, we find the following problem is of special interest.

**Problem 4.** *Determine whether there exists an algorithm (a polynomial-time algorithm) which, given an integer $n \geq 1$ and a rational number $\varepsilon > 0$, computes a rational number x such that $|x - \mathrm{pw}(K_n)| < \varepsilon$.*

# 8 Acknowledgements

We are very grateful to Daria Schymura, Carsten Thomassen, and David Wood for their valuable comments, encouragement and helpful discussions.

Marcin Kamiński gratefully acknowledges support from the Actions de Recherche Concertées (ARC) fund of the Communauté française de Belgique and from the Fonds National de la Recherche Scientifique (F.R.S. – FNRS).

Paul Medvedev gratefully acknowledges the generous support of the International NRW Graduate School in Bioinformatics and Genome Research and the AG Genominformatik group at Bielefeld, as well as the German Academic Exchange Service (DAAD) Research Grant for funding his stay in Bielefeld.

Martin Milanič gratefully acknowledges the support by the group "Combinatorial Search Algorithms in Bioinformatics" funded by the Sofja Kovalevskaja Award 2004 of the Alexander von Humboldt Stiftung and the German Federal Ministry of Research and Education.